# Studies on the Software Testing Profession

Luiz Fernando Capretz
Electrical & Computer Eng.
Western University
London, Canada
lcapretz@uwo.ca

Pradeep Waychal
CRICPE
Western Michigan University
Kalamazoo, USA
pradeep.waychal@gmail.com

Jingdong Jia
School of Software
Beihang University
Beijing, China
jiajingdong@buaa.edu.cn

Daniel Varona  Yadira Lizama
Cultureplex Laboratories
Western University
London, Canada
{dvarona, ylizama}@uwo.ca

*Abstract*—This paper attempts to understand motivators and de-motivators that influence the decisions of software professionals to take up and sustain software testing careers across four different countries, i.e. Canada, China, Cuba, and India. The research question can be framed as "How many software professionals across different geographies are keen to take up testing careers, and what are the reasons for their choices?" Towards that, we developed a cross-sectional but simple survey-based instrument. In this study we investigated how software testers perceived and valued what they do and their environmental settings. The study pointed out the importance of visualizing software testing activities as a set of human-dependent tasks and emphasized the need for research that examines critically individual assessments of software testers about software testing activities. This investigation can help global industry leaders to understand the impact of work-related factors on the motivation of testing professionals, as well as inform and support management and leadership in this context.

## I. INTRODUCTION

Despite the importance of the software industry, only a handful of studies have been done on motivating software engineers to take up and sustain testing careers. Considering the importance of software testing to the development of high quality, reliable software systems, and inadequate empirical evidence about the human factors affecting this activity, we decided to conduct a survey to try to find out what and how work-related factors motivate or demotivate software professionals follow a software testing career path?

Weyuker et al. [1] observed that the most skilled software testers were accustomed to changing jobs in their companies and becoming programmers, analysts, or system architects, because a career in software testing was not considered advantageous enough for most of the professionals. Santos et al. [2], Deak et al. [3], Florea and Stray [4], Waychal and Capretz [5], and Fernández-Sanz et al. [6] have also investigated the reasons for the lack of interest of software professionals in testing careers, and speculated on a broad range of motives.

## II. METHODOLOGY

A questionnaire was designed to collect responses on the motivation of software professionals to work as software testers, and to understand work-related factors in the specific context of software testing. Specifically, we asked professionals for the probability that they would choose testing careers by offering multiple choices: "Certainly Yes," "Yes," "Maybe," "No," and "Certainly Not". Besides, we asked the respondents to provide an open-ended but prioritized list of PROs and CONs, and open-ended rationale regarding their decisions on taking up testing careers.

Our sample consisted of 220 software professionals from four different countries (22 from India, 20 from Canada, 34 from China, and 144 from Cuba). While the Indian responses were sought from professionals, who were attending a testing conference, the Canadian responses were sought from alumni of a software engineering program at a university. The Chinese responses came from professionals, who were doing part time courses at a university. The Cuban professionals worked as software developers and taught/studied at a university. We, thus, used convenience sampling both in terms of the countries as well as the professionals. Due to a varying number of respondents in the four geographies, we use percentage instead of absolute number of responses increasing the validity of comparisons.

### A. Chances of software engineering professionals taking up testing careers

The percentage chances are depicted in Table I below.

TABLE I. PROBABILITY OF TAKING TESTING CAREER

| Responses | Canada | China | Cuba | India |
|---|---|---|---|---|
| Certain Not | 15% | 3% | 17% | n.a. |
| No | 30% | 23% | 47% | n.a. |
| May be | 30% | 59% | 15% | n.a. |
| Yes | 10% | 12% | 16% | n.a. |
| Certain Yes | 15% | 3% | 6% | n.a. |

In the four geographic regions surveyed, we found that testing was not a popular career option among software professionals. Canada has the highest percentage of professionals (25%) who wanted to take up testing careers. Many Chinese professionals were ambivalent and that was perhaps due to a relatively lower unemployment rates. Most of the Cuban professionals were very much against taking up testing careers, which may be due to better employment prospects as software developers.

### B. PROs of testing careers as perceived by professionals

The responses from each country were analyzed and presented in Table II below.

TABLE II. PERCENTAGE OF MOTIVATION DRIVERS FOR SOFTWARE TESTING PROFESSIONALS

| Motivators | Canada | China | Cuba | India |
|---|---|---|---|---|
| Learning Opportunities | 34% | 36% | 45% | 28% |
| Important job | 16% | 7% | | 28% |
| Thinking job | 7% | | | 37% |
| Easy job | 16% | 32% | | |
| More jobs | 13% | 14% | | |
| More monetary benefits | 5% | 5% | | |
| Suitable for "freshers" | | | 16% | |
| Proximity to customers | | | 16% | |
| Good infrastructure | | | 5% | |
| Increase product quality | | | 13% | |

*C. CONs of testing careers as perceived by professionals*

The survey respondents from each country were analyzed and displayed in Table III below.

TABLE III. PERCENTAGE OF DE-MOTIVATION DRIVERS FOR SOFTWARE TESTING PROFESSIONALS

| De-motivators | Canada | China | Cuba | India |
|---|---|---|---|---|
| Second-class citizen | 24% | 7% | 15% | 46% |
| Career development | 22% | 15% | 7% | |
| Complexity | 10% | 27% | 20% | 40% |
| Tedious | 17% | 25% | | 6% |
| Missed development | 12% | 9% | | 6% |
| Less monetary benefits | 10% | 9% | 13% | |
| Finding other's mistakes | | | 23% | |
| Detail oriented skills | | | 17% | |

### III. DISCUSSIONS

Testing offers tremendous learning opportunities as reported by professionals across the four countries. Barring Indian professionals, whose most voted PRO for testing being thinking jobs, professionals from the other three countries voted that as the most common PRO. Indians professionals voted that as the second PRO. The Chinese professionals' second PRO, on the other hand, was easiness of jobs. Except Cubans, other professionals also viewed the importance of testing jobs as another PRO. The Cuban PROs were, barring learning opportunities, different and included the suitability of testing jobs for inducting "freshers", proximity to customers, and an increase in commitment to software quality. We need to further investigate the reasons for such differences from the Cuban contingent.

The most common de-motivators appeared to be the second-class citizen treatment meted out to the testers and complexity resulting in stress and frustration. Except Indian professionals, others have concerns about career development and monetary benefits in testing tracks, and barring Cuban professionals, others were concerned about tediousness and missing development aspects of testing careers. Cuban professionals had different views and pointed out difficulties in finding mistakes of others, requirement of detail oriented skills as CONs of testing career. In fact, 'finding mistakes' was the most voted CON by the Cuban professionals.

### IV. CONCLUSIONS

The general empirical findings on motivation to take up and continue with testing careers suggest that testing jobs remain unattractive across the four countries. The treatment of testing professionals as second-class citizen and complexities resulting in stressful and frustrating situations appear to be common de-motivators. Our discovery of motivators and de-motivators for software testers can help global testing managers and team leaders who are dealing with motivational problems in prospective and current professionals in software testing. As previously emphasized [7], software testing is a human-dependent activity and the motivation of software testers can impact the quality of the final product and help guide industrial practitioners to handle testing personnel issues using soft skills [8].

We have to recognize that the human factors are affected by psychological and sociological factors [9], and other environmental factors such as the organizational structure and internal policies and processes [10]. Practicing managers may be able to attract larger numbers of professionals to testing careers and retain the current testing professionals by understanding common, as well as country specific motivational and de-motivational factors. Finally, the testing profession seems to be changing with the advent of Agile methods, DevOps and other paradigms. For instance, Developer in Test is a new role in many companies and this requires other competences than that of a traditional tester from decades ago. Similarly, test automation, security testing, etc, are different forms of this profession. These aspects should be taken into account in the future.